# Transition from the $C_3$–dominated discharge to the sooting plasma

Shoaib Ahmad[1,2,*], Bashir Ahmad[1], and Tasneem Riffat[1]
[1]*Accelerator Laboratory, PINSTECH, P. O. Nilore, Islamabad, Pakistan*
[2]*National Centre for Physics, Quaid-i-Azam University Campus, Shahdara Valley, Islamabad, 44000, Pakistan*

[*]Email: sahmad.ncp@gmail.com

## Abstract

Mass spectrometry and photoemission spectroscopy of a graphite hollow cathode source identify the parameters of the transition from the $C_3$-dominated discharge to the sooting plasma. The transition is a function of the shape and profile of a special cusp magnetic field $B_z(r,\theta)$, the geometry of the source, the discharge current, and pressure. Characteristic atomic and molecular emission lines and bands in the $C_3$ discharge transform into broad bands emitted by the excited soot. We identify four prominent emission bands between 300– 400 nm to be the hallmark of the sooting plasma.

## INTRODUCTION

When a cylindrical graphite hollow cathode source is operated in the glow discharge mode with Ne or Ar, atomic and molecular carbon species ($C_1$, $C_2$, $C_3$, ...) are efficiently sputtered from the cathode. However, the initiation and sustenance of the discharge critically depend on the shape and intensity of the three-dimensional (3D) cusp magnetic field $B_z(r,\theta)$ generated by permanent magnets arranged in hexapole geometry and wrapped around the cylindrical source. Such a carbon cluster source is described in detail elsewhere [1]. By manipulating the geometry of the hollow cathode (HC), the hollow anode (HA) and the region of confinement of the cusp field $B_z(r,\theta)$, we reported the soot formation properties of this source [2]. The axial $B_z$, radial $B_r$ and the azimuthal $B_\theta$ field contours produce the 3D magnetic field $B_z(r,\theta)$ that can be made to extend over the desired region of the plasma-cathode wall interactions. In this paper we provide evidence of the transition from a $C_3$–dominated discharge to the sooting plasma and highlight the parameters of this transition. Our experimental indicators are the mass spectrometry of the charged carbon clusters with a specially designed $E \times B$ velocity filter [3] and the photoemission spectroscopy of the source to determine the state of the carbon vapor. The source's geometrical and physical parameters allow us to achieve either a $C_3$–dominated discharge or a sooting plasma that may contain the entire range of the linear chains, rings and possibly, the closed caged carbon

clusters—the fullerenes [2]. The most common feature of the carbon cluster formation from graphite by sputtering [1], la- ser ablation [4], and the high pressure arc discharge [5] involves the generation of the C bond breaking sequences from the graphite surface followed by the re-bonding processes in carbon vapor. Investigations into the dynamics of the cluster formation in the hollow cathode source reveal that the cathode wall sputtering transforms the initial $Ne^+$-dominated discharge into a carbonaceous one. In such a multi-component discharge all of its constituents recycle and regenerate the cathode deposited C clusters by kinetic ($E(Ne^+,C_m^+) \sim$ 500–1000 eV) and potential ($E(Ne^*,C^*) \sim$ 10–20 eV) sputtering. However, the potential sputtering by the metastable $Ne^*$ is shown to be the most efficient regenerative agent for the clusters on the cathode surface [6].

Our present investigation into the role of the sputtering of the graphite hollow cathode by the plasma constituents yields useful information on certain similar aspects of the plasma wall sputtering in tokomak reactors. Whereas, we have designed the source to effectively enhance the wall erosion due to the plasma-wall interactions, the reverse is the requirement for the fusion reactors. Our conclusions are also valid for the studies of such processes in tokomaks. Our results show that the radial component of the extended 3D cusp magnetic field contours create an ideal environment for the entrapment of the charged particles. The subsequent interactions of the trapped, positively charged species with the cathode walls provide an efficient recycling mechanism for the wall deposited material. Therefore, we are purposefully relying on a particular source design that is to be avoided by the fusion reactor designers. In addition to providing an in-sight into the transition from a $C_3$–dominated discharge to the sooting plasma, we have presented the photo emissive signatures of a multi-component carbon plasma and its regenerative behavior.

## EXPERIMENT AND RESULTS

Schematic diagram of the source is shown in Fig. 1 with the discharge being generated and confined within the annular region between HC and HA. This is achieved by; (a) producing a threaded surface on the inside of HC so that a sharply peaked edge faces the anode, and (b) adjusting the cusp field $B_z(r, \theta)$ in the centre of HA by specially designed soft iron rings described in Ref. [1]. Discharge species are extracted from the aperture in HC by a conical extractor (not shown in the figure). Three spectra are shown at different discharge currents $i_{dis}$. Figure 1(a) is at $i_{dis}$=150 mA, Fig. 1(b) at 37.5 mA and Fig. 1(c) is at 12.5 mA. All other experimental parameters remain the same. The peak labelled $C_3^+$ is the main feature while the $Ne^+$ shifts to lower energy in Fig. 1(b) and disappears in Fig. 1(c). In the insets are the enlarged peaks of $C_1^+$ and $C_2^+$ with their respective magnification factors. The other variable intensity features, besides the sharply peaked Ne$^+$, are broad humps labelled as $C_3^{2+}$ and $C_3^{4+}$. The respective spectra in Figs. 1(a)–1(c) are obtained at gradually decreasing $i_{dis}$.

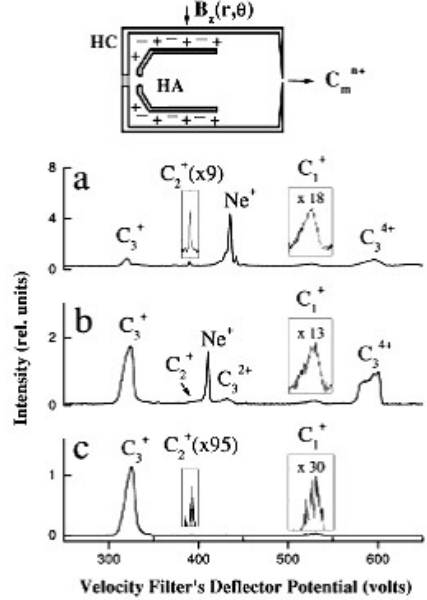

FIG. 1. The schematic diagram of the source is shown with the discharge being generated between the HC and HA. The cusp magnetic field $B_z(r,0)$ is confined in the center of HA. Charged carbon clusters $C_m^{n+}$ ($m,n \geq 1$) are extracted by a conical extractor (not shown in the figure). (a) $i_{dis}$=150 mA, Ne$^+$ is the most significant peak followed by $C_3^+$ and $C_3^{4+}$. **Inset:** $C_2^+$ and $C_1^+$ are shown magnified by a factor of 9 and 18, respectively. (b) $i_{dis}$=37.5 mA, $C_3^+$ is the most prominent mass peak with Ne$^+$, $C_3^{2+}$, and $C_3^{4+}$. $C_2^+$ is only indicated by an arrow. **Inset:** $C_1^+$ is 13 times enlarged. (c) $i_{dis}$=12.5 mA, $C_3^+$ is the only and most significant peak. Inset: $C_2^+$ and $C_1^+$ are shown by their respective enlargement factors.

Therefore, the most convincing proof of the discharge being dominated by $C_3^+$ comes from Fig. 1(c) that is obtained at $i_{dis}$ =12.5 mA ( $P_{dis}$= 8 W). We cannot ignite the source at such low power but once a stable discharge is established, the source can operate at substantially reduced $i_{dis}$. But this can only happen if one has a multi-component plasma and at least one of the species has high metastable energy levels; in this case it is Ne* ($E_{Ne*}$=16.7 eV). The pattern of sputtering and formation of the excited and ionized $C_1^{*,+}$ and $C_2^*$ from this source has been reported as a function of the discharge parameters [6]. These species become the agents for the formation of the C cluster layers on the cathode. Our experimental results suggest that the subsequent recycling and regeneration of these layers result in the formation of the $C_3^+$ dominated discharge as is shown in Fig. 1(c).

The role of the kinetic and potential sputtering has been evaluated in the regeneration of the soot [6]. One can either vary the $i_{dis}$ at constant gas pressures to manipulate the number of ionized and excited species or vary the support gas pressure $P_{Ne}$ while keeping $I_{dis}$ constant. This provides a powerful tool to investigate the mechanisms of cluster formation and dissociation in a carbonaceous environment. However, the mass spectrometry of the charged clusters at very high pressures becomes difficult due to vacuum problems. We use photoemission spectroscopy of the discharge at such high pressures and relate the information available from the characteristic lines and band emission from the excited atomic, ionic, and molecular species to that obtained by the mass analysis. Figure 2 shows two such spectra at different source pressures but with the same $i_{dis}$=75 mA. Pressure is increased by more than two orders of magnitude from $P_{Ne}$= 0.06 mbar in Fig. 2(a) to 20 mbar in Fig. 2(b).

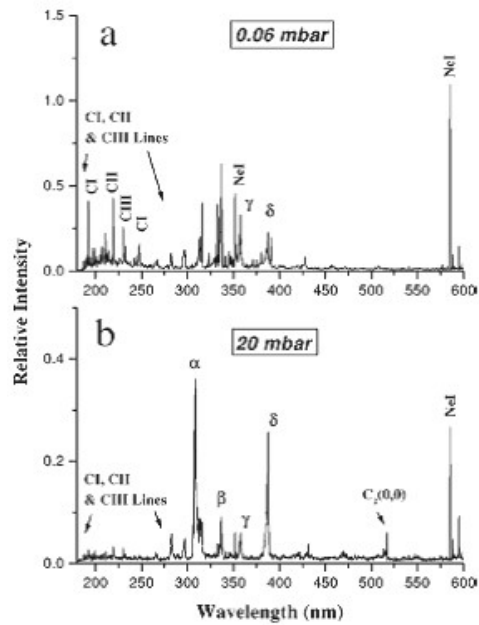

FIG. 2. (a) Photoemission spectrum at $i_{dis}$=75 mA, $P_{Ne}$ =0.06 mbar. The atomic and ionic lines are sharply defined and indicated as C I, Ne I, C II, and C III. Also shown are the two band structures at 357 nm and 385 nm are designated as $\gamma$ and $\delta$, respectively. (b) Photoemission spectrum at $i_{dis}$=75 mA and $P_{Ne}$ =20 mbar shows the greatly reduced intensities of the atomic and ionic lines of C and Ne. The four bands labelled as $\alpha$, $\beta$, $\gamma$ and $\delta$ are the prominent feature of this emission spectrum. The Swan band head of $C_2$ [ $C_2(0,0)$] at $\lambda$=5165 Å is also present with enhanced intensity.

The relative line intensities of the emission lines of C, Ne, and four bands identified as $\alpha$, $\beta$, $\gamma$ and $\delta$ are shown between 180– 600 nm. Atomic lines are sharp while the individual lines of the bands cannot be separated by our monochromator. Most conspicuous is the resonant line of Ne I at $\lambda$=5852 Å in Fig. 2(a) and 2(b). A set of lines belonging to the excited atomic $C_1^*$ i.e., C I at $\lambda$ =1931 Å and 2478 Å, the inter-combination multiplet of the singly charged $C_1^+$ i.e., C II at $\lambda$ =2324−2328 Å and the doubly charged $C_1^{++}$ i.e., C III at $\lambda$ =2297 Å are shown in the range 180–300 nm. Between 300 and 400 nm, we get a large number of prominent lines and bands, the origin of which we will discuss here and also in a later section while dealing with the photoemission spectra of the sooted plasmas. The vibrationally excited Swan band head of $C_2(0,0)$ at $\lambda = 5165$ Å is the other most important feature at higher pressures. It is present at low pressures as well but acquires significant proportions as the pressure is increased.

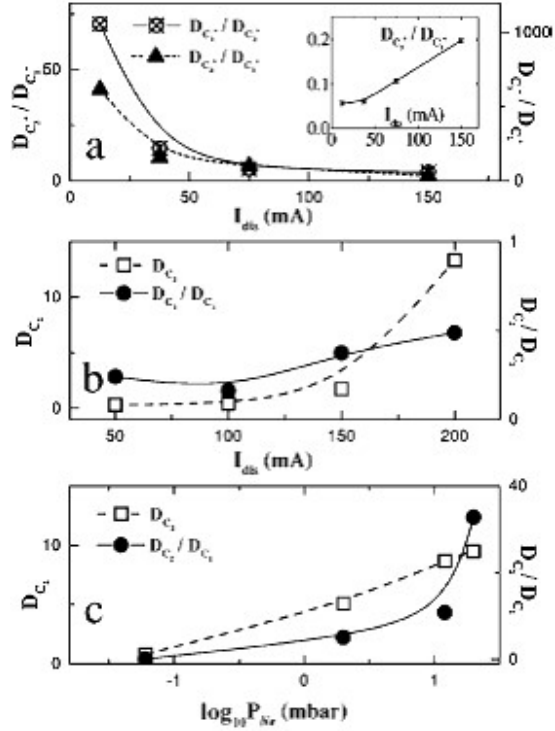

FIG. 3. (a) The ratios of the ion densities $D_{C_3^+}/D_{C_1^+}$ and $D_{C_3^+}/D_{C_2^+}$ are plotted as a function of $i_{dis}$. $C_3^+$ is the main plasma constituent and its contribution increases at lower $i_{dis}$. **Inset:** The ratio $D_{C_2^+}/D_{C_1^+}$ is shown for the same range of $i_{dis}$. (b) At $P_{Ne}$=0.06 mbar the photoemission spectra yield the number densities of the lowest vibrationally excited transition of $C_2$'s Swan band [ $C_2(0,0)$] at $\lambda$=5165 Å and its ratio with the excited atomic C atoms (C I) at $\lambda$=2478 Å denoted as $D_{C_2}/D_{C_1}$ is plotted as a function of $i_{dis}$. Same as FIG. 3(b) but with $i_{dis}$=75 mA and the discharge variable being $P_{Ne}$ between 0.06 and 20 mbar. The $D_{C_2}/D_{C_1}$ ratio rises steeply after $P_{Ne}$>1 mbar.

The accumulated data for the relative ion densities of $C_3^+$ ($D_{C_3^+}$), $C_2^+$ ($D_{C_2^+}$), and $C_1^+$ ($D_{C_1^+}$) as functions of $i_{dis}$ and $P_{Ne}$ are presented in Fig. 3. In Fig. 3(a) the ratios of $D_{C_3^+}/D_{C_1^+}$ and $D_{C_3^+}/D_{C_2^+}$ are plotted as a function of $i_{dis}$ from 12.5 to 150 mA. $C_3^+$ is the sole survivor at very low discharge currents. Its relative number density increases with respect to that of $C_1^+$ by a factor of 27 between $i_{dis}$ =150 and 12.5 mA. Whereas, $D_{C_3^+}/D_{C_2^+}$ increases by two orders of magnitude as $i_{dis}$ is decreased in the same range. This pattern is clearly evident from Fig. 1(c). In the inset of Fig. 3(a) $D_{C_3^+}/D_{C_1^+}$ is shown for the same range of $i_{dis}$. The number density of $C_2^+$ ($D_{C_2^+}$) increases from 5% to 20% of $C_1^+$ ($D_{C_1^+}$). Figures 3(b) and 3(c) show the number densities of the excited states of $C_2$ and $C_1$ as functions of $i_{dis}$ and $P_{Ne}$, respectively. We have derived the number densities of the excited $C_1^*$ and $C_2^*$ from the line intensities of the electronically excited C I at λ=2478 Å and the lowest transition of the Swan band with the vibrationally excited $C_2$(0,0) at λ=5165 Å. The number densities of $C_2$(0,0) denoted as $D$ are plotted on the left vertical axis and the ratio $D_{C_2^+}/D_{C_1^+}$ along the right vertical axis for $i_{dis}$ between 50 and 200 mA. It can immediately be seen that the density of $C_2$(0,0) is an increasing function of both of the discharge parameters $i_{dis}$ and $P_{Ne}$. The number density of the excited state of neutral $C_1$ i.e., C I also increases rapidly with $i_{dis}$, but its contribution is reduced at higher pressures as can be seen in Fig. 2(b). That is why the $D_{C_2^+}/D_{C_1^+}$ ratio in Fig. 3(b) does not vary as steeply as it does in Fig. 3(c). From the photoemission data the ratio of the singly charged to the excited C i.e., $D_{CII}/D_{CI}$ remains constant, for example, in the range of $P_{Ne}$ = 0.1 to 1 mbar. It is 0.55±0.2 for $i_{dis}$ = 50-200 mA. Similarly, $D_{C_2(0,0)}/D_{CI}$ = 2.2±0.4 under the same conditions. At low pressure discharge i.e., $P_{Ne}$ ≤ 0.1 mbar the singly charged atomic carbon C1+ (CII) and the excited diatomic molecular carbon $C_2$(0,0) are directly related with CI i.e., $C_1$ in the level $^1P_1$ ($E^1P_1$=7.5 eV). Thus it may be deduced that the origin of $C_1$ and $C_2$ is in the dissociation of $C_3$ via $C_3$—$C_1$+$C_2$. From the tabulated data presented in Fig. 3 we conclude that $C_3^+$ is not only the significant species at low $i_{dis}$ but it is the main constituent of the discharge under the experimental conditions as elucidated in Fig. 1. It may be due to the regeneration pattern of all clusters $C_m^+$ (m>3) up to $C_{30}$. This process favors the accumulation of $C_3$ as the end product [7-9].

This fragmentation scheme $C_m^+ \rightarrow C_{m-3}^+ + C_3$ (dissociation energy ~5.5±0.5 $eV$) has been predicted in the *ab initio* calculations [7] of $C_m^+$ up to m=10 and the experiments for all $C_m^+$ up to m≤60 [8,9]. $C_3$ itself can fragment via $C_3 \rightarrow C_1+C_2$ and $C_2 \rightarrow 2C_1$. This fragmentation pattern can explain not only the preponderance of $C_3$ but also the enhanced contribution of $C_1$ at higher $i_{dis}$. But at higher $P_{Ne}$, the large relative increase in the density of $C_2$ cannot be explained by the $C_3$ fragmentation. At $P_{Ne}$>1 mbar, the increased contribution from the $C_2$(0,0) is accompanied by the corresponding increase in the bands emission in 300–400 nm and a consequent decrease in the excited and ionic states of the atomic carbon $C_1$ (C I, C II, C III) lines as seen in Fig. 2(b). We interpret it as the onset of the formation of the closed caged clusters $C_m$ (m>30). These clusters further fragment via $C_2$ emission $C_m(m \geq 30) \rightarrow C_{m-2}+ C_2$ [4]. This is our preferred interpretation of the enhanced $C_2$(0,0) intensities at $P_{Ne}$>10 mbar.

The schematic diagram in Fig. 4 shows the hollow cathode source operating in the sooting mode. The magnetic cusp field provides conditions of discharge in the main HC region and is no longer confined to the annular region between HA and HC as was done to operate the source in the $C_3$-dominated mode shown in Fig. 1. A typical cluster spectrum is shown from this source configuration in Fig. 4(a) that is taken from Ref. [2].

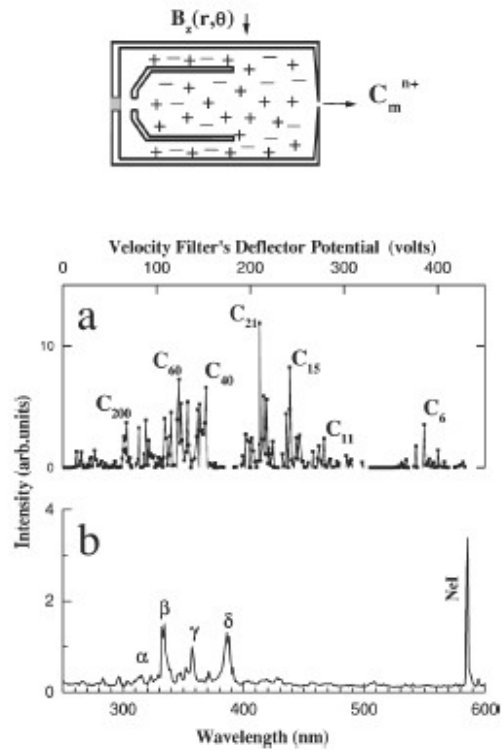

FIG. 4. The source is shown in the sooting mode with the cusp magnetic field $B_z(r,\theta)$ providing discharge conditions in the main HC. (a) A typical mass spectrum of the clusters $C_m^{n+}$ is shown. Beyond $C_6^+$, no significant cluster ions are seen. (b) The photoemission spectrum from the sooted source operating with the same discharge parameters as in (a). Besides the ubiquitous Ne I $\lambda$=5852 Å line, the significant emissions are from (most probably from $N_2$) the bands $\alpha$, $\beta$, $\gamma$ and $\delta$ at 310, 337, 357, and 385 nm, respectively.

It shows clusters labelled from $C_1$ to $C_{200}$, however, larger clusters are also present in the spectrum. The sooting mode shown here may be related to the regenerative processes initiated by the plasma−sooted cathode interactions. As pointed out in the $C_3$−dominated discharge, the regeneration of the soot layers takes place mostly by the excited and metastable plasma species like Ne* and $C_1$* with $E_{Ne}$*=16.7 eV and $E_C$*=7.5 eV, respectively. Figure 4(b) shows the photoemission spectra taken with the four broad and prominent bands designated as $\alpha$, $\beta$, $\gamma$, $\delta$ appearing at almost the same wavelengths as in Fig. 2(b). The band heads are at 310, 337, 357, and 385 nm, respectively. These are typical $N_2$ bands and their presence may provide the excited molecules with high potential energies needed to recycle the soot-covered cathode [6].

# CONCLUSION

In conclusion we have presented in this paper, the parameters for the initiation of a $C_3$−dominated discharge in a graphite hollow cathode that operates in a special cusp magnetic field configuration. We have shown that the transition from the $C_3$ to the sooted plasma is dependent upon the source geometry, the field configuration, and the discharge parameters. Mass spectra provide clear evidence of the dominant plasma species while operating either in the $C_3$−dominated regime or the sooting mode. The photoemission spectroscopy yields the characteristic emissive pattern in the form of four UV bands in the range 300– 400 nm associated with the dominant carbon cluster species. Our tentative interpretation of these bands in the light of the emission spectra shown in Ref. [6] and our present mass spectra shown in Figs. 1 and 4(a), respectively, seem to be associated with the onset of the formation of the closed caged clusters, i.e., the fullerenes and perhaps the multi-shelled carbon onions [2].

We have discussed the formation of the soot by the overall process of recycling and introduction of the cathode deposited carbon clusters. Once the $C_3 \rightarrow$ soot transition has been made, an entire range of clusters is produced in the regenerative sooting plasma. The other equally interesting question is regarding the onset of the formation of the multi-shelled carbon onions. In addition, this sooting plasma source can inject a particular carbon clusters $C_m^+ (m > 2)$ into the beam line of any accelerator for various basic and applied studies with the carbon clusters. As we have also indicated the essential features of a soot forming discharge in magnetized plasmas contained within the graphite cathodes, this information may be of interest to the fusion reactor designers.